# Fabrication of magnetocaloric La(Fe,Si)$_{13}$ thick films

N. H. Dung[1], N. B. Doan[1], P. De Rango[1], L. Ranno[1], K. G. Sandeman[2,3, a)] and N. M. Dempsey[1, a)]

[1]Univ. Grenoble Alpes, CNRS, Grenoble INP, Institut Néel, 38000 Grenoble, France
[2]Department of Physics, Brooklyn College of the City University of New York, 2900 Bedford Avenue, Brooklyn, New York 11210, USA
[3]Physics Program, The Graduate Center, CUNY, 365 Fifth Avenue, New York, New York 10016, USA

La(Fe,Si)$_{13}$–based compounds are considered to be very promising magnetocaloric materials for magnetic refrigeration applications. Many studies have focused on this material family but only in bulk form. In this paper we report on the fabrication of thick films of La(Fe,Si)$_{13}$, both with and without post-hydriding. These films exhibit magnetic and structural properties comparable to bulk materials. We also observe that the ferromagnetic phase transition has a negative thermal hysteresis, a phenomenon not previously found in this material but which may have its origins in the availability of a strain energy reservoir, as in the cases of other materials in which negative thermal hysteresis has been found. Here, it appears that the substrate acts to store strain energy. Our exploratory study demonstrates the viability of thick films of the La(Fe,Si)$_{13}$ phase and motivates further work in the area while showing that additional perspectives can be gained from reducing the dimensionality of magnetocaloric materials in which the magneto-volume effect is large.

___________________________________________

[a)] Authors to whom correspondence should be addressed: karlsandeman@brooklyn.cuny.edu and nora.dempsey@neel.cnrs.fr

## I. INTRODUCTION

La(Fe,Si)$_{13}$–based compounds displaying a giant magnetocaloric effect (MCE) are of particular interest because of their great potential for magnetic refrigeration applications.[1,2,3,4,5] The giant MCE in La(Fe,Si)$_{13}$–based alloys stems from the so-called magneto-volume effect, *i.e.* the magnetic phase transition from the paramagnetic to the ferromagnetic state is associated with an increase in volume without breaking the cubic lattice symmetry. A strong magneto-volume effect can lead to a sharp change in magnetization and in volume at the Curie temperature ($T_C$) in a first-order phase transition. However, the magneto-volume effect is also present in a phase transition of second order. In this case, the magnetic phase transition and the volume change take place gradually rather than sharply.[6] By changing composition, substituting elements or introducing interstitials, the magneto-volume effect and the MCE can be modified.[7,8]

LaFe$_{13-x}$Si$_x$ alloys are stabilized in the NaZn$_{13}$-type cubic structure in the range $1.2 < x < 2.5$. With increasing Si content, a changeover from first-order ($x < 1.6$) to second-order ($x > 1.6$) transition, along with an increase in $T_C$ from 175 to 255 K, was reported.[7,9] Interestingly, the first-order transition in LaFe$_{13-x}$Si$_x$ alloys shows only a small thermal hysteresis, making them very promising for refrigeration applications.[10,11] However, a low $T_C$ of ~ 200 K is an obstacle to applications near room temperature. Many efforts have been made to raise $T_C$, such as introducing interstitial hydrogen atoms[12] or replacing some Fe with Co.[13] While the giant MCE still remains after inserting hydrogen atoms, the substitution of Co for Fe considerably weakens the MCE since the first-order transition is gradually transformed into a transition of second order.

Studies to date have focused on La(Fe,Si)$_{13}$-based bulk alloys, with no reports on La(Fe,Si)$_{13}$ in film form. Thin or thick films may serve as model systems for materials studies and also have potential for use in micro-systems. Here we report on the fabrication of La(Fe,Si)$_{13}$ thick films that exhibit similar properties to bulk materials. Our aim is to demonstrate the viability of thick films of this compound for the first time, while motivating future work to control and optimize the nature of the phase transition. We present an unusual thermal irreversibility of magnetization, and a previously unreported negative thermal hysteresis (NTH), which appears to be a result of strain energy stored in the substrate. Finally, we demonstrate that hydrogen can be inserted into the film so as to increase the transition temperature.

## II. EXPERIMENTAL SECTION

A La-Fe-Si alloy target was prepared by induction melting from pieces of La, Fe and Si (atomic ratio La:Fe:Si = 1.2:11.5:1.5). La-Fe-Si films of thickness up to 5.3 μm were deposited by triode sputtering onto thermally oxidized Si(001) wafers at a rate of 3.8 μmh$^{-1}$. All films had 90 nm thick Ta buffer and capping layers. The films were deposited at $T_{dep}$ ~ 530 K (no power to the heater of the substrate holder, but substrate heated by the plasma) and then annealed at high temperature for 5 min under secondary vacuum in a Rapid Thermal Processing (RTP) furnace. Annealing leads to the formation of cracks in the films. This is tentatively attributed to the difference in thermal contraction upon cooling between the metallic film and the Si substrate, which may result in residual tensile stress in the annealed film at room temperature. Hydrogen absorption by annealed films was conducted at 2 bar and 623 K for 2 h. Note that no obvious additional cracking, nor film peel-off, was observed following hydrogenation. A Vibrating Sample Magnetometer (VSM) with Superconducting Quantum Interference Device (SQUID) technology (Quantum Design) was employed to characterize the magnetic properties. The structural properties, the microstructure and phase composition of the films were characterized using x-ray diffraction (XRD) with Co-K$_α$ radiation (λ = 1.789 Å) with the scattering vector perpendicular to the film surface (theta – 2 theta geometry) and field emission scanning electron microscopy (FE-SEM) with integrated energy dispersive x-ray spectroscopy (EDS).

## III. RESULTS AND DISCUSSION

The as-deposited films exhibit very broad XRD peaks corresponding to $α$-Fe (not shown). Figure 1 shows room-temperature XRD patterns for 5.3 μm thick La-Fe-Si films after post-deposition heat treatment at 1073 and 1173 K. It can be seen that these thick La-Fe-Si films crystallize in the NaZn$_{13}$-type cubic structure. The relative intensities of the peaks indicate that the films are polycrystalline with randomly oriented grains, i.e., no preferential texture. Additional diffraction peaks are attributed to $\alpha$-Fe, a La-rich phase, $Fe_7Ta_3$ and Fe-containing $\beta$Ta. The formation of the latter two phases due to diffusion between the main layer and the buffer and capping layers during the annealing process is supported by a literature report on the binary Ta-Fe systems.[14] No $NaZn_{13}$-type phase was detected after annealing 5.3 μm thick films below 973 K (data not shown). Note that the $NaZn_{13}$-type phase was only detected in films of thickness greater than 2 μm, and that the volume content of this phase was maximum in the 5.3 μm thick films. Hereafter we will only present results for these films. The eventual role that stress plays in stabilizing the $NaZn_{13}$-type phase will be the subject of a future study.

Backscattered electron SEM images of fractured cross sections of 5.3 μm thick La-Fe-Si films before and after heat treatment are shown in the inset of Fig. 1. The very small bright areas in the images of the annealed samples are attributed to the La-rich phase identified by XRD. Moreover, the increase in the thickness of the buffer and capping layers, and the apparent formation of two distinct sub-layers in the annealed samples, support our conclusion concerning diffusion from the main layer into the buffer and capping layers to form two distinct Fe-Ta phases.

Considering the 5.3 μm thick La-Fe-Si films before and after annealing at 1173 K, EDS analysis conducted on the as-deposited and annealed films indicate compositions of $LaFe_{11.2}Si_{1.9}$ and $LaFe_{10.9}Si_{2.1}$, respectively. This change in composition may be attributed to the diffusion of Fe into the Ta layers. Rietveld analysis of all the peaks on the diffraction pattern of the annealed film associated with the $La(Fe,Si)_{13}$ phase (data not shown) gives a lattice parameter value of a = 11.4469(24) Å. According to lattice parameter values reported for bulk alloys by Shen et al.[7], this corresponds to a Si content of 2.5(1), which is larger than the value of 2.1 that we estimated by EDS. Comparison of the Curie temperature of the annealed film (see below) and those reported by Shen et al.[7], also points to a Si content close to 2.5. The apparent difference in Si content may be explained by the relatively high error associated with EDS analysis, in particular when characterizing films. Note that the theta – 2 theta XRD geometry used here does not allow us to probe anisotropic strain, which could be expected due to high stress in thick films. A detailed analysis of the influence of strain on XRD diffraction patterns is beyond the scope of this work, but will be addressed in a future study.

Figure 2a shows magnetization as a function of temperature in a field of 0.01 and 0.1 T measured in the in-plane direction for the 5.3 μm thick La-Fe-Si film after annealing at 1173 K. A broad transition from the paramagnetic to the ferromagnetic state might suggest either a second-order magnetic phase transition with $T_C$ = 245 K, or a series of first order transitions with slightly varying values of $T_C$, the highest being around 245 K. While no information could be found concerning the magnetic properties of $Fe_7Ta_3$ or Fe-containing $\beta$Ta, $Fe_2Ta$ is reported to be a Pauli paramagnet.[15] The spontaneous magnetization above the Curie transition of the main La-Fe-Si phase is thus attributed to the α-Fe phase identified in XRD patterns. This corresponds to roughly 10% of the low temperature magnetization, and thus α-Fe would account for close to 5 vol% of the film, as its spontaneous magnetization is roughly twice that of $La(FeSi)_{13}$. The spontaneous magnetization of the La-Fe-Si phase at 5 K is thus estimated to be of the order of $9 \times 10^5$ $Am^{-1}$. In-plane and out-of-plane *M-H* curves measured at 5 K and 300 K are compared in Fig. 2b. The out-of-plane *M-H* curve at 5 K shows hard axis behavior, saturating at a field of 1.18 T, or 0.94 MA/m. Such a value can be explained by shape anisotropy, considering a demagnetizing factor of 1 for films measured in the out-of-plane direction. The potential contribution of magnetocrystalline anisotropy is considered to negligible compared to shape anisotropy, as the $La(FeSi)_{13}$ phase is cubic in the ferromagnetic and paramagnetic states and the grains are randomly oriented. Possible contributions from magnetoelastic anisotropy induced either by the differential thermal expansion of the film and the substrate or a magnetovolume effect around $T_C$ may exist but are not needed to explain the observed anisotropy. At 300 K, when the LaFeSi is paramagnetic, the M-H curves are much less anisotropic, indicating that α-Fe is present as inclusions which are almost spherical in shape (demagnetization factor ∼ 1/3). Using a Maxwell relation[3] for the magnetization isotherms near $T_C$ (see Fig. 2b), the isothermal entropy change is estimated to be about 60 kJm$^{-3}$K$^{-1}$ for a 7 T field change (see Fig. 2c). These values are comparable to that of bulk alloy with similar composition.[1,7,9]

Thermal hysteresis is a signature of a first-order phase transition, and it is not expected to occur in materials showing a second-order magnetic phase transition.[16] We clearly observe a separation of the magnetization curves on cooling and heating in the vicinity of $T_C$ (see Fig. 2a), which indicates that the magnetic phase transition in our films is not truly second-order. However, close inspection reveals that the $T_C$ measured on heating is smaller than that measured on cooling. It should be noted that the $M$-$T$ data were recorded in sweep mode at various rates of temperature change from 1 to 5 K/min. No considerable change of the M-T curves was observed due to different heating or cooling rates, indicating that the observed negative thermal hysteresis is not an instrumental artifact caused by thermal lag.

This negative thermal hysteresis (NTH) observed here is opposite to the thermal hysteresis normally observed in materials with a first-order ferro – paramagnetic phase transition, which normally exhibit a larger $T_C$ on heating than that on cooling.[10,17] However, NTH does not run contrary to the laws of thermodynamics if the magnetic phase probed by the VSM constitutes only one part of the total material system, and if the other parts of the material system are coupled to the magnetic phase via a strain energy term that is significant. Recently, the thermal hysteresis of FeRh thin films was reduced by coupling the films to a substrate; the elastic hysteresis was merely transferred to the substrate.[18]

In general, it is known that such substrate effects can reduce the thermal hysteresis of a first-order phase transition, and even switch its sign to that of NTH. For this reason, NTH has been observed in multi-phase bulk materials and films and multi-layer materials on substrates, where strain energy has a significant bearing on both the temperature of a first order transition and its hysteresis. Examples include calorimetry measurements on Ca$_4$[Al$_6$O$_{12}$]SO$_4$ (with -31 K of NTH),[19] and Sr$_4$[Al$_6$O$_{12}$]SO$_4$ (with -5 K of NTH)[20] and resistance measurement on Ti-Ni-W thin films (with about -3K of NTH).[21] These aforementioned alloys fall into a category of so-called thermoelastic shape memory alloys in which stored strain energy can assist in the progress of a transition. Early examples of thermoelastic shape memory, such as beta-brass exhibit an austenite start temperature on heating which can be higher than the martensite start temperature on cooling[22] but the overall heating and cooling curves do not have a complete reversal.

In the case considered in our study, we posit that the magnetovolume effect across the transition results in a strain coupling to the substrate that varies greatly with temperature across the phase transition, thus straining the substrate in addition to the magnetically active film. Strain energy build up in the substrate can be released via the magnetoelastic transition of the La-Fe-Si phase, allowing the latter to exhibit NTH. The key to this understanding is the notion that the magnetic signal is not a complete indicator of the energetic state of the whole "sample" (including substrate). Our findings are a logical extension of the finding of reduced thermal hysteresis in thin films of FeRh[18] but are the first observation of NTH in a thick film magnetocaloric material.

From Fig. 2a, we can also see a second thermomagnetic phenomenon: a weak bump in magnetization apparent upon cooling in a low field of 0.01 T, extending to below 100 K. We note that such a phenomenon was not observed in a higher field of 0.1 T (see inset of Fig. 2a). This irreversibility may originate from the thermal evolution of strain in the La-Fe-Si layer caused by the neighboring layers since the La(Fe,Si)$_{13}$ materials display a negative thermal expansion in the vicinity of $T_C$.[6,11] A bump in magnetization was also observed when melt-spun La(Fe,Si)$_{13}$ hydrides were cooled under hydrostatic pressure in a low magnetic field, which supports the idea that strain may cause the effect. [10,17] The presence of cracks in our film is evidence of relaxation of in-plane tensile stress, and a residual tensile stress may exist in the film at room temperature. This tensile stress may be expected to decrease in the ferromagnetic regime due to magnetovolume expansion when cooling through $T_C$ and the Invar effect at lower temperatures. Passage through a stress-free state may explain the observed magnetization bump. An alternative source of explanation may be the relative alignment of the film and the magnetic field, or spin glass or magnetic frustration effects. These arguments are only speculative and more work is required to establish the origins of this weak feature in the lowest measurement fields.

Comparing the *M-T* curves measured in the in-plane configuration for the 1073 K annealed sample before and after hydrogenation (see Fig. 3), it can be seen that $T_C$ is increased by approximately 100 K upon hydrogenation. Hydrogen absorption may occur through the capping layer, as Ta is highly permeable to hydrogen. [23] However, cracks formed during the post-deposition annealing step, that can be clearly observed in plan-view SEM images (see insets of Fig. 3), are expected to serve as fast routes for hydrogen entry into the La-Fe-Si layer. Indeed, in light of the presence of these cracks, the long-term stability of hydrogenated film will be studied. The magnetic transition is found to be sharper in hydrogenated films. Finally, the unusual phenomena described above persist after hydrogenation.

## IV. CONCLUSIONS

In summary, the preparation of La(Fe,Si)$_{13}$ in thick-film form has been reported for the first time. While films show values of Curie temperature, lattice constant, field-induced isothermal entropy change and spontaneous magnetization comparable with values reported for Si-rich bulk material, they also display unusual features, *i.e.* both a negative hysteresis, the first of its kind for a magnetocaloric thick film, and the appearance of a weak magnetization bump upon cooling. Quantification of the origin of these effects requires further investigation and modeling and we hope that this study will promote further activity in the area of strained, first-order magnetocaloric films. Finally, we have shown that a La(Fe,Si)$_{13}$ thick film can be hydrogenated so as to increase the transition temperature to above room temperature. The successful fabrication of La(Fe,Si)$_{13}$ thick films suggests prospects for the study of new multi-phase material properties and for the detailed manipulation and exploitation of the ferromagnetic phase transition for use in, for example, micro-systems applications.

**Acknowledgements**

The research leading to these results has been partially funded by the 7th Framework Program of the European Commission under the grant agreement No. 310748 (DRREAM). This work has also benefited from the support of the project HiPerTherMag ANR-18-CE05-0019 of the French National Research Agency (ANR). The authors are grateful to Dominique Givord, Martino Lo Bue, Julia Lyubina and Shihua Zhao for fruitful discussions and to Richard Haettel (Institut Néel) and Isabelle Gelard (formerly at CNRS-CRETA) for technical assistance. KGS would like to thank the Université Grenoble Alpes for sponsoring his visiting researcher position during 2017.

**Data Availability Statement**

The data that support the findings of this study are available from the corresponding authors upon reasonable request.

**References**

[1] K. Gschneidner Jr, V.K. Pecharsky, and O. Tsokol, Reports Prog. Phys. **68**, 1479 (2005).

[2] K.G. Sandeman, Scr. Mater. **67**, 566 (2012).

[3] E. Brück, J. Phys. D. Appl. Phys. **38**, R381 (2005).

[4] V. Franco, J.S. Blázquez, B. Ingale, and A. Conde, Annu. Rev. Mater. Res. **42**, 305 (2012).

[5] A. Smith, C.R.H. Bahl, R. Bjørk, K. Engelbrecht, K.K. Nielsen, and N. Pryds, Adv. Energy Mater. **2**, 1288 (2012).

[6] R. Huang, Y. Liu, W. Fan, J. Tan, F. Xiao, L. Qian, and L. Li, J. Am. Chem. Soc. **135**, 11469 (2013).

[7] B.G. Shen, J.R. Sun, F.X. Hu, H.W. Zhang, and Z.H. Cheng, Adv. Mater. **21**, 4545 (2009).

[8] L.F. Bao, F.X. Hu, L. Chen, J. Wang, J.R. Sun, and B.G. Shen, Appl. Phys. Lett. **101**, 162406 (2012).

[9] K. Niitsu, S. Fujieda, A. Fujita, and R. Kainuma, J. Alloys Compd. **578**, 220 (2013).

[10] J. Lyubina, R. Schäfer, N. Martin, L. Schultz, and O. Gutfleisch, Adv. Mater. **22**, 3735 (2010).

[11] F. Hu, B. Shen, J. Sun, Z. Cheng, G. Rao, and X. Zhang, Appl. Phys. Lett. **78**, 3675 (2001).

[12] A. Fujita, S. Fujieda, Y. Hasegawa, and K. Fukamichi, Phys. Rev. B **67**, 104416 (2003).

[13] B. Rosendahl Hansen, L. Theil Kuhn, C.R.H. Bahl, M. Lundberg, C. Ancona-Torres, and M. Katter, J. Magn. Magn. Mater. **322**, 3447 (2010).

[14] V.T. Witusiewicz, A.A. Bondar, U. Hecht, V.M. Voblikov, O.S. Fomichov, V.M. Petyukh, and S. Rex, Intermetallics **19**, 1059 (2011).

[15] H.P.J. Wijn, editor, *Magnetic Properties of Metals: D-Elements, Alloys and Compounds* (Spinger, Berlin, 1991).

[16] A. Tishin and Y. Spichkin, *The Magnetocaloric Effect and Its Applications IOP* (Institute of Physics Publishing, Bristol and Philadenphia, 2003).

[17] J. Lyubina, K. Nenkov, L. Schultz, and O. Gutfleisch, Phys. Rev. Lett. **101**, 177203 (2008).

[18] Y. Liu, L.C. Phillips, R. Mattana, M. Bibes, A. Barthélémy, and B. Dkhil, Nat. Commun. **7**, 11614 (2016).

[19] D. Kurokawa, S. Takeda, M. Colas, T. Asaka, P. Thomas, and K. Fukuda, J. Solid State Chem. **215**, 265 (2014).

[20] H. Banno, S. Ichikawa, S. Takeda, T. Asaka, M. Colas, P. Thomas, and K. Fukuda, J. Ceram. Soc. Japan **125**, 364 (2017).

[21] P.J.S. Buenconsejo, R. Zarnetta, D. König, A. Savan, S. Thienhaus, and A. Ludwig, Adv. Funct. Mater. **21**, 113 (2011).

[22] A.L. Titchener and M.B. Bever, J. Met. **6**, 303 (1954).

[23] K.S. Rothenberger, B.H. Howard, R.P. Killmeyer, A. V Cugini, R.M. Enick, F. Bustamante, M. V Ciocco, B.D. Morreale, and R.E. Buxbaum, J. Memb. Sci. **218**, 19 (2003).

**Figure captions**

Fig. 1. Room temperature XRD patterns of 5.3 µm thick La-Fe-Si films after annealing at 1073 K and 1173 K, measured in theta – 2 theta geometry with the scattering vector perpendicular to the film surface. The inset shows backscattered electron SEM images of fractured cross-sections of the La-Fe-Si films before (left) and after annealing at 1073 (middle) and 1173 K (right).

Fig. 2. Temperature-dependent magnetization on cooling and heating measured in the in-plane direction in magnetic fields of 0.01 T and 0.1 T (inset) (a), the magnetization as a function of magnetic field in the in-plane and out-of-plane directions (b), and the isothermal entropy change under field changes in the vicinity of $T_C$ (c) of 5.3 µm thick La-Fe-Si films after annealing at 1173 K.

Fig. 3. Temperature-dependent magnetization on cooling and heating in an in-plane magnetic field of 0.01 T for a 1073 K annealed 5.3 µm thick La-Fe-Si films before and after hydrogenation. Insets show a top-view image of the sample that is acquired with an in-lens detector and schematic demonstration of the hydrogenation process.

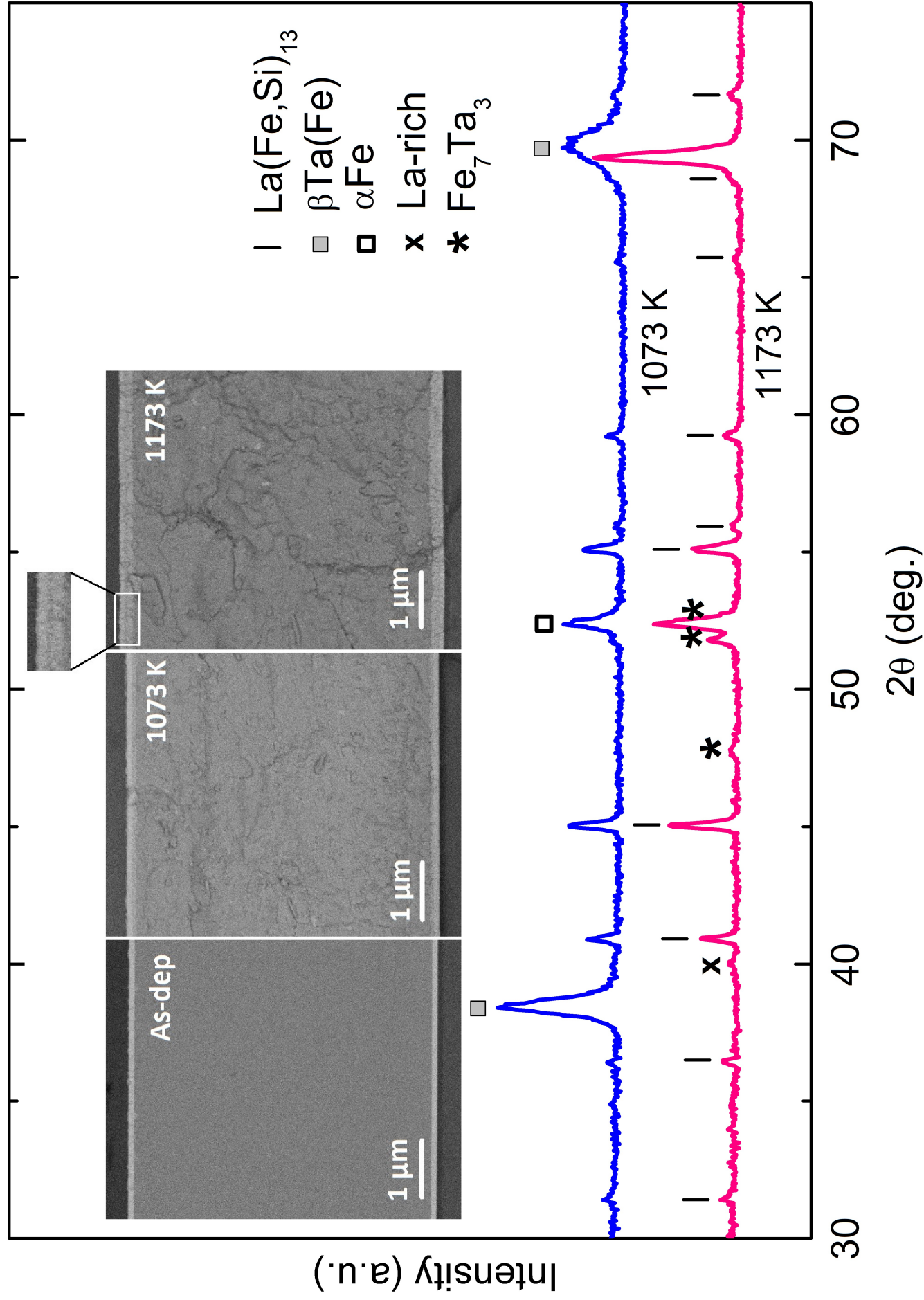

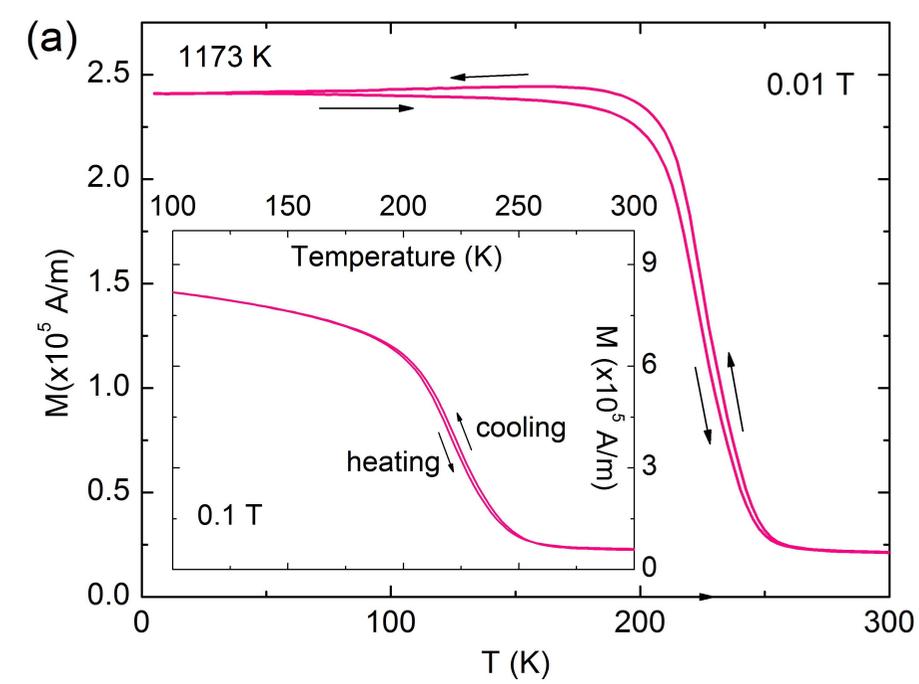

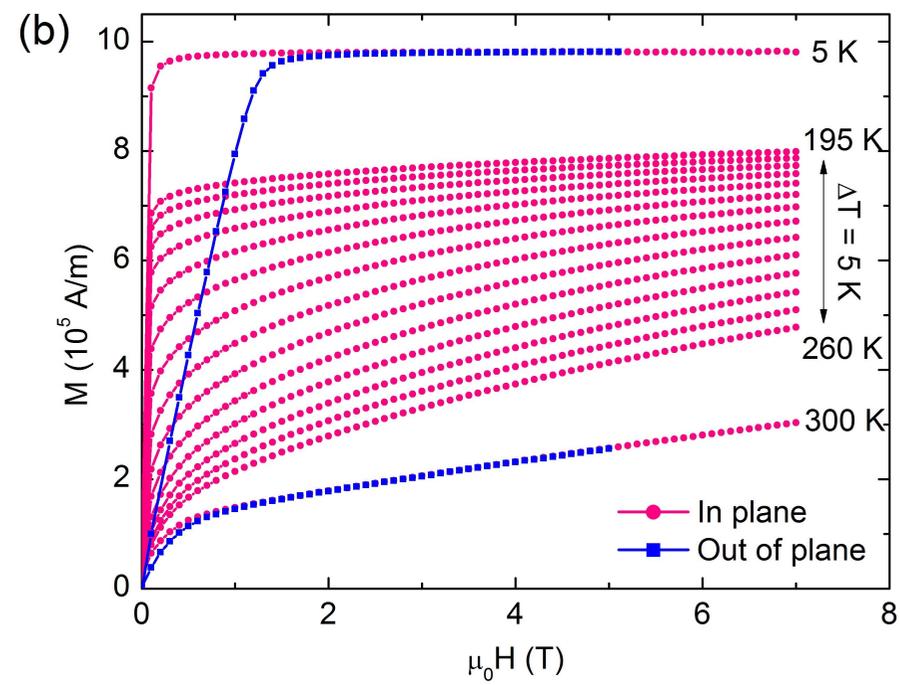

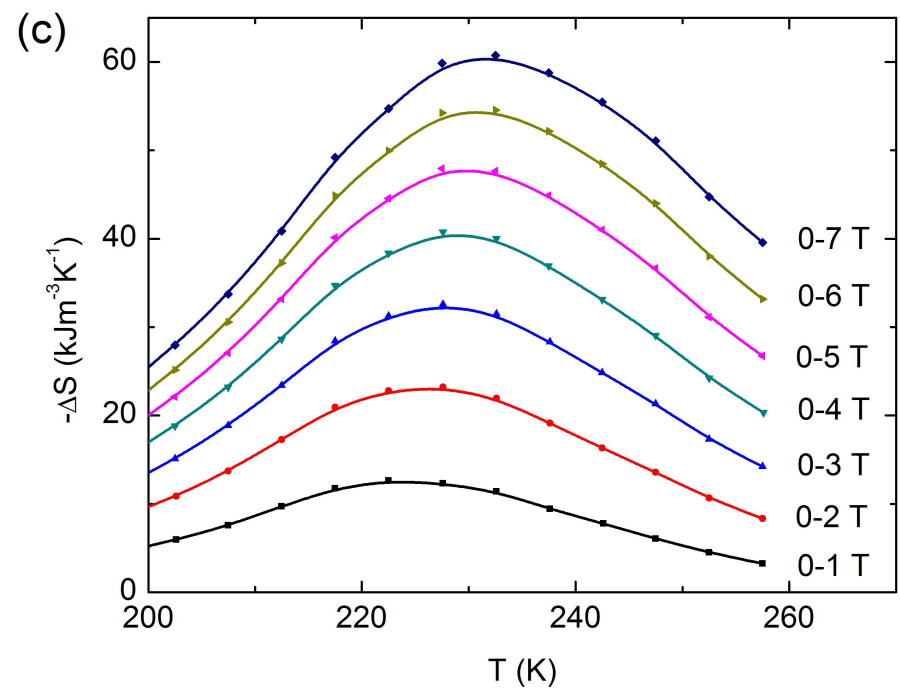

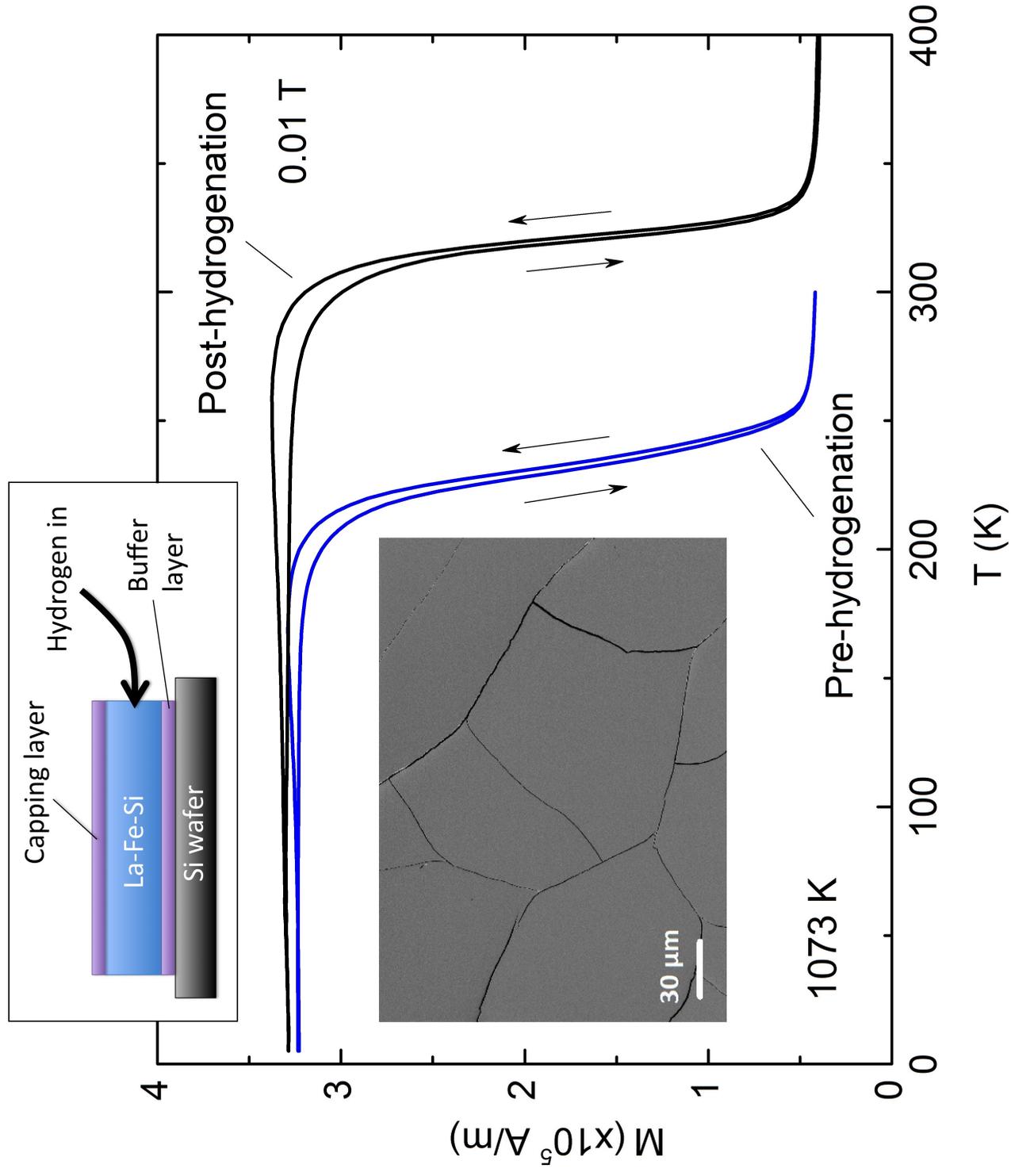